\documentclass[onecolumn]{IEEEtran}
\usepackage{amsmath}
\usepackage{setspace}
\usepackage{amsfonts}
\usepackage{textcomp}
\usepackage{graphicx}
\usepackage{color}
\usepackage{float}
\usepackage{subfig}

\renewcommand{\d}[2]{\frac{d#1}{d#2}}

\newcommand{\avg}[1]{\langle #1 \rangle}

\DeclareMathOperator*{\argmax}{argmax}

\newcommand{\st}[1]{#1}
\renewcommand{\st}[1]{}
\newcommand{\logg}{\text{log}_2}
\graphicspath{{./figure/}}
\begin{document}
\doublespacing
\title{Enhancing classical target detection performance using nonclassical Light}\author{Han Liu, Amr S.~Helmy,~\IEEEmembership{Fellow,~OSA,}~\IEEEmembership{Senior Member,~IEEE,}
}
\markboth{Magazine of xxx , ~Vol.~xx, No.~x, July~2019}%
{Haoyu \MakeLowercase{\textit{et al.}}: Non-classical Semiconductor Photon Sources Enhancing the Performance of Classical Target Detection Systems}
\maketitle
\begin{abstract}
In this article, we demonstrate theoretically and experimentally how one can exploit correlations generated in monolithic semiconductor quantum light sources to enhance the performance of optical target detection. A prototype target detection protocol, the quantum time-correlation (QTC) detection protocol, with spontaneous parametric down-converted photon-pair sources, is discussed. The QTC protocol only requires time-resolved photon-counting detection, which is phase-insensitive and therefore suitable for optical target detection. As a comparison to the QTC detection protocol, we also consider a classical phase-insensitive target detection protocol based on intensity detection. We formulated the target detection problem as a probe light transmission estimation problem, and we quantify the target detection performance with the Fisher information criterion and the receiver operation characteristic analysis. Unlike classical target detection and ranging protocols, the probe photons in our QTC detection protocol are completely indistinguishable from the background noise and therefore useful for covert ranging applications. Finally, our technological platform is highly scalable and tunable and thus amenable to large scale integration necessary for practical applications.
\end{abstract}

\section{INTRODUCTION}
Optical target detection has been receiving increasing attention owing to many emerging applications in the domains of computing, human/machine interaction, LIDAR, and non-invasive biological imaging, amongst others. Conventionally, the sensitivity of optical target detection could be improved by increasing the source brightness, detector sensitivity or improving the throughput of the optical setup. In addition, computational imaging has been providing astounding advantages to the fidelity of various imaging modalities  \cite{gariepy2016detection}, those can be utilized on top of any hardware improvement brought about due to enhanced device performance.  Recent work has also explored the extension of target detection techniques and equipment to novel illumination wavelengths suitable for remote sensing of distant objects  \cite{barzanjeh2015microwave}. Practical platforms for these purposes will find applications in sensing, surveillance, and autonomous driving.

Aside from the aforementioned approaches, a novel and radical way to improve the target detection performance is to use a non-classical state of light as the source.  
Current LIDAR systems are based on illumination sources that can be described by classical electromagnetism. However, the properties of classical states of electromagnetic radiation pose ultimate limits, such as the thermal and shot noise.
On the other hand, non-classical states of electromagnetic radiation could exploit two key properties arising in quantum mechanics, namely \emph{quantization of the electromagnetic field} (where the quantized excitations are referred to as \emph{photons}) and  \emph{quantum entanglement and correlation}.
\footnote{Correlation means a degree of freedom of two particles always takes random but correlated values.  Quantum entanglement between two particles means that the two particles are correlated in different degrees of freedoms in such a way that it is not possible to fully describe the state of the two particles individually.}
It has been shown that the quantum entanglement  \cite{lloyd2008enhanced} and correlations \cite{Lopaeva:2013,england2019quantum} that exist within non-classical states of light could be utilized to enhance the accuracy of target detection in the lossy and noisy environment.\\

The basic idea of non-classical source enhanced target detection pivots upon non-classical \emph{twin beams} in either radio or optical frequencies. The nonclassical twin beams constitute the \emph{probe} and \emph{reference} light (also referred in the literature as \emph{signal} and \emph{idler} light, respectively) that are highly entangled and correlated with each other.
The probe light is sent towards the target and the back-reflected light is collected for detection.
The reference light is stored and detected locally.
Through analyzing the quantum entanglement, or classical correlation, between the collected light (which could be probe light or noise light when the target is absent) and the reference light with an appropriate measurement scheme, it is possible to distinguish the presence and absence of the target object even in the presence of strong environmental noise.
This is because, unlike the probe light, environmental noise is not correlated or entangled with the reference light. 

One of the most prominent approaches to utilize such quantum advantages is \emph{quantum illumination}, which utilizes the entanglement between the reference and probe light to significantly boost the target detection sensitivity beyond the classical limit \cite{lloyd2008enhanced}.
Recent work suggests that such benefits of entanglement could even survive through an extremely lossy and noisy target detection channel  \cite{Zhang:2015}.
However, in order to profit from such entanglement benefit, a significant level of complexity in the instrumentation, including phase-sensitive joint detection\footnote{a joint measurement requires bringing the two photons to the same physical location at the same time.}, is essential. For example, the implementation in  \cite{Zhang:2015} requires sub-wavelength-level stabilization of optical phases between the probe and reference light, which is far from practical if the target distance is unknown or fluctuating. As such, formidable challenges lie ahead on the route to harvesting the entanglement advantages. An alternative strategy to enhance the target detection performance while mitigating the complexity of quantum illumination is to only use classical correlations that exist in the non-classical twin beams. Unlike entanglement, the measurement of correlation only entails independent measurements of the probe and reference light and is therefore much more feasible for practical implementation.  The key aspect of correlation enhanced target detection protocols is the type of correlation that is utilized, which could be intensity correlation, amplitude correlation or time-frequency correlation, etc.\\ 

The most important component of correlation enhanced target detection protocols is the non-classical light sources. There has been astounding progress in the prowess of such non-classical sources in the last decade  \cite{Kang:2016, Kang:2015}. In particular, it has been shown that integrated monolithic semiconductor devices can be used to generate and tailor high-quality quantum states of light in active semiconductor structures, such as aluminum gallium arsenide (AlGaAs) devices  \cite{Valles:2013, Horn:2012}. Such structures can directly produce non-classical twin beams without any additional off-chip interferometry, spectral filtering, compensation, or post-selection. The generated twin beams could be effectively coupled into optical fiber or integrated target detection systems. The flexibility in waveguide structure design also allows for efficient dispersion control and quantum state engineering.\\

In this article, we provide a review of the recent advances in optical target detection that exploit the strong temporal correlation of non-classical twin beams \cite{liu2019enhancing,trans}. We theoretically analyze and experimentally demonstrate a prototype target detection protocol with a semiconductor waveguide source. The experimental result shows substantial enhancement of the target detection performance as compared to the classical protocol that does not utilize correlation.
Moreover, we show that the semiconductor waveguide source is suitable for practical implementations due to its flexible tailorability and potential in high-density integration.

\section{CORRELATION ENHANCED TARGET DETECTION}
\begin{figure*}[t]
\centering
\includegraphics[width=0.8\columnwidth]{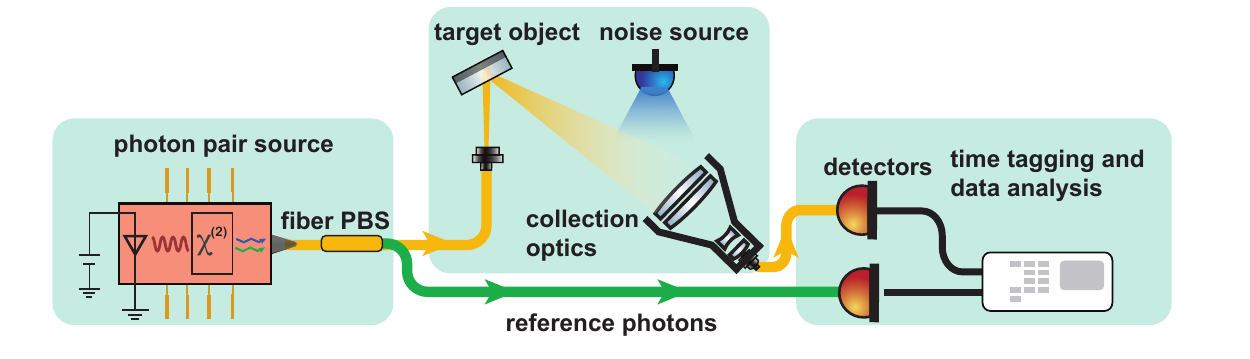}
\caption{The schematic of general correlation enhanced target detection protocols. Left: the non-classical photon pair source; middle: the free-space transceiver of the probe photons; right: the detection of the probe and the reference photons and the data analysis.} \label{SETUP}
\end{figure*}

The general schematic of correlation enhanced target detection protocols consists of three sections: \emph{source}, \emph{transceiver}, and \emph{detection}, as shown in Fig. \ref{SETUP}.
In the source section, the non-classical probe and reference light are generated and separated (through, for example, wavelength or polarization demultiplexing). The probe and reference light could be (simultaneously) correlated in many degree of freedoms, such as  such as time, frequency and intensity, etc. The probe light is sent to probe the target in the transceiver section. The back-reflected light from the target object (none if the target object is absent) is collected for detection in the detection section. Despite the presence of the target object, strong background noise always couples into the detection section and gets detected. The background noise is assumed to be continuous-wave (CW) and broadband white noise that is completely uncorrelated with the reference light. 
The reference light is directly sent to the detection section and get detected. The optical path lengths of the reference and probe light do not have to be balanced.
The detection section consists of two photo-detectors (the probe detector and reference detector) that can resolve the correlated degree of freedoms of the probe and reference light. The presence or absence of the target object could be determined by analyzing the correlation between the collected light and the reference light since only probe light that is reflected by the target object could be correlated with the reference light. The \emph{correlation enhancement} of the target detection performance is most pronounced in a noisy environment: the strong correlation between the probe and reference light can help to easily distinguish the reflected probe light from the noise background. The measurement of the correlation only requires independent and separate measurements of the probe and reference light. Therefore the non-classical entanglement between them is not directly observed. However, the correlation enhancement of the target detection performance can still be considered as a quantum benefit because the correlation strength between the probe and reference light can exceed the classical limit\cite{Lopaeva:2013,england2019quantum}. For comparison, the current state-of-the-art in LIDAR simply transmits photons towards the target and observes the reflection. In other words, no correlation of any kind is used. This classical protocol will be referred to as the CI (classical intensity) protocol in the rest of the paper. In the case of radars, it is possible to do better than this using match filtering. However, that is not easily possible in the optical regime. Hence, such a direct intensity detection scheme for LIDAR can be considered as the baseline for comparison. \\

The key aspect of correlation enhanced target detection protocols is the correlation between the probe and reference light. To date, the most widely adopted approach to generate correlated twin beams is spontaneous parametric down conversion (SPDC). The generated probe and reference light could be correlated in multiple degrees of freedom, based on which different types of target detection protocols could be designed. To give an intuitive and quantitative comparison between different protocols, it is instructive to quantify the strength of different types of correlation. Assuming the detection outcome of the two detectors obey the joint distribution \(p(x,y)\), the correlation between the probe light and the reference light could be quantified as the \emph{mutual information} (MI) \cite{bavaud2005introduction}:  
\begin{gather}
\text{MI} = \sum\limits_{x\in \mathcal{X},y\in\mathcal{Y}} p(x,y)\text{log}_2(\frac{p(x,y)}{p(x)p(y)})\label{MIEQ}
\end{gather}
where \(\mathcal{X},\mathcal{Y}\) are the spaces of detection outcome \(x,y\) for the probe and reference detector. Since the target detection performance is directly limited by the transmitted probe power, it is useful to consider the normalized mutual information, which is defined as the mutual information divided by the average number of transmitted probe photons. The expression of the normalized mutual information could serve as a theoretical measure for the correlation enhancement of the target detection performance, despite that its specific expression may not be directly used for practical performance characterization.
\subsection{Intensity Correlation}
The intensity correlation that exists within the probe and reference light simply means that the number of generated probe photons and reference photons are always equal. In other words, the detection of a reference photon implies that one single probe photon has been sent to probe the target object. Intensity correlation has already been used in quantum imaging to reduce the noise below the standard quantum limit \cite{brida2010experimental}. In the implementations \cite{Lopaeva:2013,england2019quantum,He:2018} of intensity correlation enhanced target detection protocols (will be referred to as intensity correlation protocols in the rest of this paper), the correlated twin beams are generated through pulse pumped SPDC: a short wavelength pump light pulse with narrow temporal duration interacts with the nonlinear medium and generates a pair of short duration wave packets for the probe and reference light. The number of generated probe and reference photons in their respective wave packets are always equal. The detection section of the intensity correlation protocols consists of time-resolved photon-counting detectors for both the collected light and reference light. The temporal resolution of the detectors is required so that each probe light wave packet and its correlated reference light wave packet are resolved and ``paired'' correctly. In practical implementations, the nonlinear conversion efficiency is usually very low that the probability \(\bar{n}\) of generating a probe (reference photon) in a wave packet is much less than unity. Under this approximation, the probability distribution \(p_I(x,y)\) of generating \(x\) probe and \(y\) reference photons is given by:  
\begin{gather}
p_I(0,0) = 1-\bar{n}\hspace{1cm}p_I(1,1) = \bar{n} \\
p_I(1,0) = p_I(0,1) = 0 
\end{gather}
The normalized mutual information \(\text{MI}_\text{I}\) between the probe and reference photon number could be calculated according to \eqref{MIEQ}:
\begin{gather}
\text{MI}_\text{I} = \frac{ -\bar{n}\logg(\bar{n}) - (1-\bar{n})\logg(1-\bar{n}) }{\bar{n}} \simeq -\logg\bar{n}
\end{gather}
As can be seen, the intensity correlation is very high with low mean probe photon number per pulse \(\bar{n}\) but diminishes rapidly as \(\bar{n}\) increases. Therefore for pulse pumped intensity correlation protocols, the correlation enhancement is limited to the low \(\bar{n}\) regime \cite{england2019quantum}. 

\subsection{Temporal Correlation}
\begin{figure}[h]
\centering
\includegraphics[width=0.6\columnwidth]{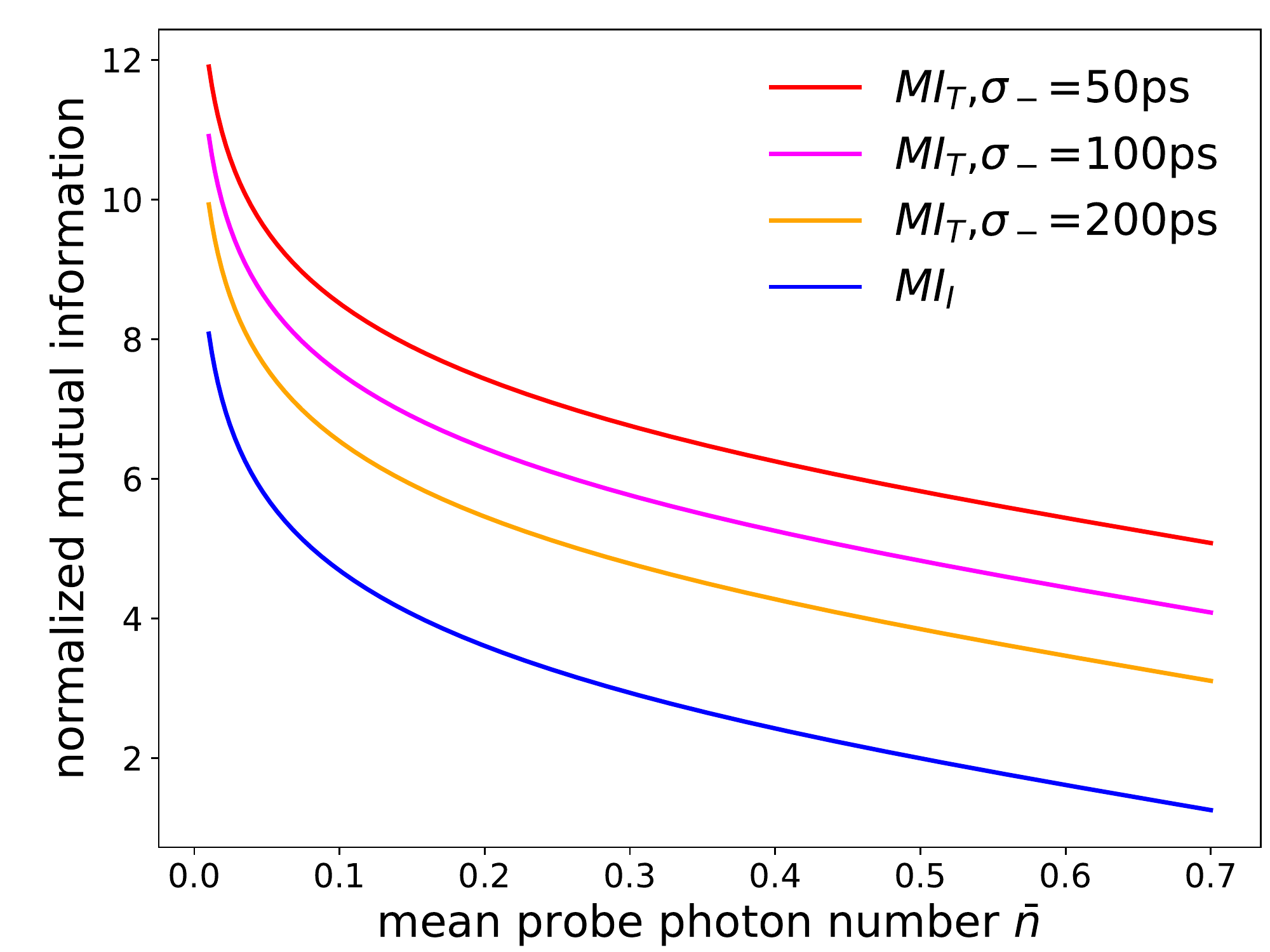}
\caption{Comparison between the normalized mutual information for target detection protocols enhanced by (blue) intensity correlation and (red,magenta and orange) temporal correlation. The pump pulse duration \(\sigma_+\) for the temporal correlation protocol is set to 1 ns.}\label{MI}
\end{figure}

For the intensity correlation protocol that is based on pulse pumped SPDC process, the correlation enhancement is limited to the low mean photon number per pulse (\(\bar{n}\)) regime. A way to increase the target detection performance without sacrificing the correlation enhancement is to increase the repetition rate of the probe pulses but keep the mean probe photon number per pulse \(\bar{n}\) fixed. In the limit where the repetition period equals the pulse duration of the pump light, the pump light becomes CW. This generalized protocol with CW pump light will be termed the quantum temporal correlation (QTC) protocol since the probe and reference photon are generated in pairs `almost simultaneously' \cite{liu2019enhancing}. The correlation time \(\Delta t\), which is defined as the standard deviation of the time difference between the generation of probe and reference photon, quantifies the intrinsic temporal correlation. To utilize such temporal correlation, the detectors must be able to resolve the detection time of each incoming photon. It is worth noting that the detection of the probe and reference photons does not have to be simultaneous since their optical path lengths are not equal in general. The temporal correlation analysis is done in post data processing. For simplicity of the following analysis, the detectors are assumed to have a perfect temporal resolution. To model the probability distribution of the different detection outcomes, it is convenient to consider a long quasi-CW pump pulse, during which the probability of generating a photon pair is \(\bar{n}\). When a photon pair is generated, then the conditional joint probability distribution of detection time \(t_p,t_r\) of the probe and reference photon, respectively, could be approximated as a bivariate Gaussian distribution. Then the joint probability \(p_T\) of the detection outcomes is:  
\begin{gather}
p_T(\text{no pairs}) = 1-\bar{n}\\
p_T(t_p,t_r) = \frac{\bar{n}}{2\pi\sigma_+\sigma_-} \exp(-\frac{(t_p-t_r)^2}{4\sigma^2_-}-\frac{(t_p+t_r)^2}{4\sigma^2_+})
\end{gather}
where \(\sigma_+\) characterizes the duration of the quasi-CW pump pulse and \(\sigma_-=\Delta t/\sqrt{2}\) quantifies the intrinsic temporal correlation time of the photon pairs. The normalized mutual information \(\text{MI}_T\) corresponds to the probability distribution of the measurement outcomes is given by:
\begin{gather}
\text{MI}_\text{T} = \text{MI}_\text{I} -\frac{1}{2}\logg(1-(\frac{\sigma_+^2-\sigma_-^2}{\sigma_+^2+\sigma_-^2})^2)
\end{gather}
The additional term of \(\text{MI}_\text{T}\) is the temporal correlation enhancement, which is inversely related to the pump duration \(\sigma_+\) and directly related to the intrinsic temporal correlation time \(\Delta t\). Ideally, to achieve large correlation enhancement, the intrinsic temporal correlation time \(\Delta t\) should be as short as possible. However, it should be noted that the detector temporal resolution could limit the amount of intrinsic temporal correlation that could be utilized for target detection. The comparison between the normalized mutual information for different protocols is shown in Fig. \ref{MI}.  

\section{FIGURE OF MERIT FOR TARGET DETECTION PROTOCOLS}
In correlation enhanced target detection protocols, multiple detectors (i.e. the probe and reference detector) are used to resolve the correlation between the probe and reference photon. Therefore the probability distribution of the different detection outcomes is multi-dimensional. However, it is generally hard to directly infer the presence or absence of the target object from the experimentally observed multidimensional distribution. Instead, a practical approach for target detection is to construct a \emph{target detection signal}, defined in terms of the experimentally measured quantities, to quantify the probability of the presence of the target object. Based on different definitions of target detection signal, the target detection performance could also be quantified by different figures of merit. A general and suitable figure of merit for target detection is important for quantifying and comparing the performances of different target detection protocols.\\
\subsection{The Signal-to-noise Ratio Figure Of Merit}
\begin{figure}
\centering
\includegraphics[width=0.6\columnwidth]{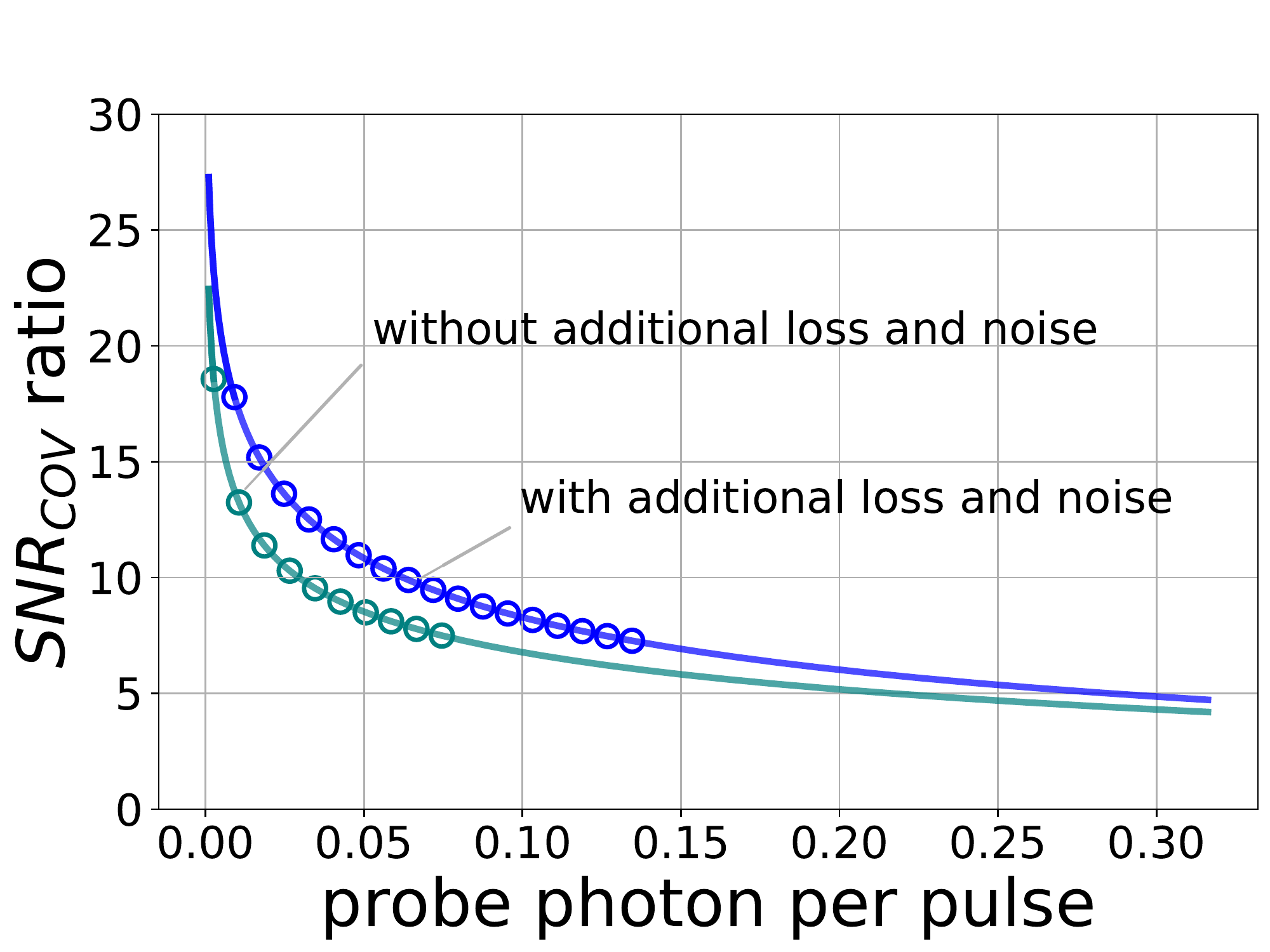} 
    \caption{The performance advantages (quantified as the ratio of the \(\text{SNR}_{COV}\)) of the intensity correlation protocol as compared to the classical intensity correlation protocol. Blue (teal) line: with (without) the additional loss and noise in the target detection channel. The detail of the experimental result and analysis could be found in  \cite{He:2018}. }\label{PowerSNR}
\end{figure}

For intensity correlation protocols, one way to construct the target detection signal is to directly define it as the covariance between the output of the probe and reference detectors. In reference  \cite{Lopaeva:2013,He:2018} the target detection signal is defined as the covariance between the number of photons that are detected on the probe and the reference detector: 
\begin{gather}
S_{COV} = N_pN_r - \avg{N_p}\avg{N_r}
\end{gather}
where \(N_p\) and \(N_r\) stand for the number of collected photons and reference photons detected in a pair of probe and reference wave packet. The angle brackets stand for statistical averages over multiple SPDC pump pulses. A unique feature of this definition is that the average value of \(\avg{S_{COV}}\) is always zero when the target object is absent. Therefore \(\avg{S_{COV}}\) is resilient to the fluctuation of environment noise power. The signal-to-noise ratio (SNR) of the target detection is quantified as the contrast of \(S_{COV}\) (normalized by its standard deviation) when the target object is present or absent: 
\begin{gather}
\text{SNR}_{COV} = \frac{\avg{S_{COV}}_{in}-\avg{S_{COV}}_{out}}{\sqrt{\avg{\Delta^2S_{COV}}_{in}+\avg{\Delta^2S_{COV}}_{out}}}
\end{gather}
where the subscript \(in\) and \(out\) stand for the presence and absence of the target object and \(\avg{\Delta^2 S_{COV}}\) is the variance of \(S_{COV}\). Note that \(\avg{S_{COV}}_{out}\) is always zero. To show the performance advantage of the intensity correlation protocol, previous works \cite{Lopaeva:2013,He:2018} considered a similar classical intensity correlation protocol that is based on a correlated thermal twin-beam source, which has been proven to be the optimal source for classical intensity correlation \cite{Lopaeva:2013}. The ratio of the performance (quantified as \(SNR_{COV}\)) of the intensity correlation protocol and the classical intensity correlation protocol is plotted in Fig. \ref{PowerSNR}, as a function of mean probe photon number per pulse for different background noise and loss condition. The detail of the experimental result could be found in  \cite{He:2018}. As can be seen, the performance advantage of the intensity correlation protocol is limited to the low mean probe photon number per pulsed regime.  \\ 

A more straightforward definition of target detection signal for intensity correlation protocols is the rate of coincidence detections (\(N_c = N_pN_r\)), which is defined as the probability of detecting photons on both the probe and reference detectors in a pair of probe and reference wave packet \cite{england2019quantum}. The definition of the target detection signal is chosen in this manner since coincidence detections directly reflect the paired-generation nature of probe and reference photons. The target detection SNR is quantified as the contrast of \(N_c\) when the target object is present or absent:        
\begin{gather}
\text{SNR}_{NC} = \frac{\avg{N_c}_{in}-\avg{N_c}_{out}}{\avg{N_c}_{out}}
\end{gather}
where, as before, the subscript \(in\) and \(out\) stand for the presence and absence of the target object and the angle brackets stand for statistical averages. It could be shown that, under this figure of merit, the correlation enhancement of the intensity correlation protocol, as compared to the CI protocol with direct intensity detection only, is given by the second-order cross-correlation function \(g^{(1,1)}\) of the probe and reference light \cite{england2019quantum}: 
\begin{gather}
g^{(1,1)}= \frac{\avg{N_pN_r}}{\avg{N_p}\avg{N_r}}
\end{gather}
Since \(g^{(1,1)}\) is inversely proportional to the number of probe and reference photons generated per pulse, it is clear that the correlation enhancement of the intensity correlation protocol is limited to the low mean probe photon number per pulse regime. This is in agreement with the previous analysis with the mutual information criterion. \\ 
\subsection{The Transmission Estimation (Fisher Information) Figure Of Merit}
Although the previously mentioned definitions of target detection signal and figure of merit demonstrate the quantum enhancement of intensity correlation protocols compared to their classical counterparts, they are implementation-dependent and artificial to some extent. For example, the definition of SNR cannot be easily generalized for the QTC protocol. Therefore, to quantify and compare the performances of different correlation enhanced target detection protocols, a more general definition of target detection signal and figure of merit is needed. The solution to this problem to formulate target detection as a transmission estimation problem: from the experimentally measured photo-detection statistics, it is possible to obtain an estimation \(\hat{\eta}_p\) of the total transmission \(\eta_p\) of the probe light \footnote{The total transmission of the probe light is defined as the ratio between the total number of probe photons (not including the noise photons) detected and the total number of probe photons generated.}.
The estimated transmission \(\hat{\eta}_p\) is a natural choice of the `target detection signal' since its average value is positive if and only if the target object is present. The variance of transmission estimation, on the other hand, characterizes the minimal value of transmission that could be distinguished from zero transmission, and therefore quantify the target detection sensitivity. According to the parameter estimation theory  \cite{ly2017tutorial}, the maximal likelihood estimation \(\hat{\eta}_{p,MLE}\) could achieve the minimal estimation variance, which is given by the inverse of the corresponding Fisher information FI: 
\begin{gather}
\hat{\eta}_{p,MLE} = \argmax\limits_{0\le \eta_p\le 1} p_{\eta_p}(x) \\
\text{FI} = \sum\limits_{x'} p_{\eta_p}(x') (\d{}{\eta_p}\text{log}p_{\eta_p}(x'))^2
\end{gather}
where \(p_{\eta_p}(x')\) is the probability distribution function of the different experiment outcomes \(x'\) and  \(x\) are the actual experiment outcome. The value of the Fisher information FI depends on many factors such as the environment loss, target object reflectivity, and the detection efficiency. \\

The transmission estimation formalism could also be used for receiver operation characteristics (ROC) analysis, which is a widely adopted figure of merit for target detection.  
When the target object is present (\(\mathbf{H_1}\)) or absent (\(\mathbf{H_0}\)), the probe transmission estimation signal \(\hat{\eta}_{p,MLE}\) could be modeled as:  
\begin{align}
\mathbf{H}_1&: \hat{\eta}_{p,MLE} =\eta_p+\mathbf{n}(\eta_p)\\ \nonumber
\mathbf{H}_0&: \hat{\eta}_{p,MLE} =\mathbf{n}(0)
\end{align}
where $\mathbf{n}(\eta_p)$ is the estimation uncertainty as a function of the probe photon transmission \(\eta_p\) (\(\eta_p=0\) when the target object is absent). The presence or absence of the target object could be decided with certain detection probability \(P_\text{d}\) and false alarm probability \(P_{fa}\) \footnote{The detection rate is defined as the probability of successfully detecting the target object when the object is actually present. The false alarm rate is defined as the probability of falsely predicting the presence of the target object while the target object is actually absent. } by comparing \(\hat{\eta}_{p,MLE}\) to a threshold value \(V\). By varying the threshold \(V\), the relationship between the detection rate \(P_\text{d}\) and the false alarm rate \(P_\text{fa}\) (widely known as the ROC curve) could be derived. By assuming that the estimation uncertainty $\mathbf{n}(\eta_p)$ is a Gaussian random variable with variance given by the inverse of the Fisher information (FI), the theoretical prediction of the ROC curve could be obtained \cite{trans}.

\subsection{The Performance Of The Temporal Correlation Enhanced Target Detection Protocol}
Transmission estimation and Fisher information criterion are useful for characterizing the QTC protocol performance. For the QTC protocol, the detection section consists of time-resolved single-photon detectors (the probe and reference detector). Since the target detection channel could be considered as stationary, i.e. target not moving, the absolute time of each photon detection event is not informative. Therefore only time information that relates the time difference of different target detection events could be of interest. Such statistics must be the number of coincidence detections \(N_c\), which is defined as detecting two photons on both the probe and reference detector, with the detection time difference less than a small coincidence window \(T_c\). Two other photon detection statistics of interest are the number of photon detection events \(N_p,N_r\) on the probe and reference detector that do not contribute to coincidence detections. It could be shown that \(N_c,N_p,N_r\) each obeys an independent Poisson distribution. The Fisher information for transmission estimation could be calculated from the joint probability distribution of \(N_p,N_c\) and \(N_r\) \cite{liu2019enhancing}:
\begin{gather}
\text{FI} = (\frac{\eta_r^2\nu^2}{P_c}+\frac{(1-\eta_r)^2\nu^2}{P_p}+\frac{\eta_r^2\nu^2}{P_r})\tau\\
P_c =  \eta_r\eta_p \nu+\eta_r\nu_b\nu T_c\\
P_r =  \eta_r \nu - P_c\hspace{20pt}\\
P_p = \nu_b+ \eta_p\nu - P_c \label{eq1}
\end{gather}
The definitions of different variables are summarized in table \ref{TAB}. The Fisher information \(\text{FI}\) increases as the coincidence window \(T_c\) decreases, suggesting that the performance of the QTC protocol is related to the temporal correlation. The minimal coincidence window is limited by the intrinsic temporal correlation \(\Delta t\) of the twin beams as well as the detector temporal resolution \cite{liu2019enhancing}. It worth noting that the CI protocol, which only consists of intensity detection on the probe detector, could be considered as a special case of the QTC protocol where the transmission of the reference photon \(\eta_r\) is set to zero.  
\begin{table}[h]
\centering
\caption{table list of variables}
\begin{tabular}{|l|l|}
\hline
\(P_c\)                   & coincidence detection rate (Hz)             \\ \hline
\(P_r\)                   & reference detector detection rate (Hz)      \\ \hline
\(P_p\)                   & probe detector detection rate (Hz)          \\ \hline
\(\nu\)                   & probe photon generation rate (Hz)           \\ \hline
\(\nu_b\)                 & number of noise photons detected per second (Hz) \\ \hline
\(\eta_p\)               & total transmission of the probe light       \\ \hline
\(\eta_r\)             & total transmission of the reference light   \\ \hline
\(T_c\)           & coincidence window                          \\ \hline
\(\tau\)         & detection time (s) \\ \hline
\end{tabular}\label{TAB}
\end{table}
\section{SYSTEM DEMONSTRATION}
To date, most of the prior work on quantum illumination and correlation enhanced target detection work has been using bulk optics. While notable in terms of demonstrating a certain type of quantum advantage, they employed intensity correlation. The correlation enhancement has been shown in terms of the ROC metric, which is fundamental in radar detection theory. However, for it to be of interest to radar practitioners, advantage has to be demonstrated in the regime where the $P_\text{fa}$ is of the order of $10^{-6}$. No prior work has demonstrated this in experimental investigations, be it optical or microwave. In this section, we provide an overview of the first such demonstration \cite{trans,liu2019enhancing} based on the QTC protocol in a setup that lends itself to be a scalable solution.
\subsection{Experimental Setup}
As shown in the schematic plot Figure. \ref{SETUP}, the QTC experimental setup consists of three sections: the non-classical photon-pair source, the free-space transceiver, and the detection system.
The photon pair source is a 5\(\mu\)m x 1mm semiconductor waveguide that is pumped by an external CW Ti-Sapphire laser at 783 nm. The broadband, CW probe (with horizontal polarization) and reference (with vertical polarization) light are separated by a polarization beam-splitter. 
The reference light is immediately sent to and detected by the reference detector in the detection section. 
The probe light is sent to the transceiver section and emitted through an optical collimator towards the target object. 
The back-scattered probe light from the target object is collected by a telescope and detected by the probe detector in the detection section.  To simulate strong environmental noise, broadband CW noise light from a light-emitting diode shines towards the collection optics, leading to a large number of noise photons getting detected on the probe detector.
 The three sections of the setup are connected via single-mode fibers, which enables a flexible system deployment.

\subsection{Semiconductor Sources Of Non-classical Photon Pairs}
\begin{figure}[h]
\centering
\includegraphics[width=0.6\columnwidth]{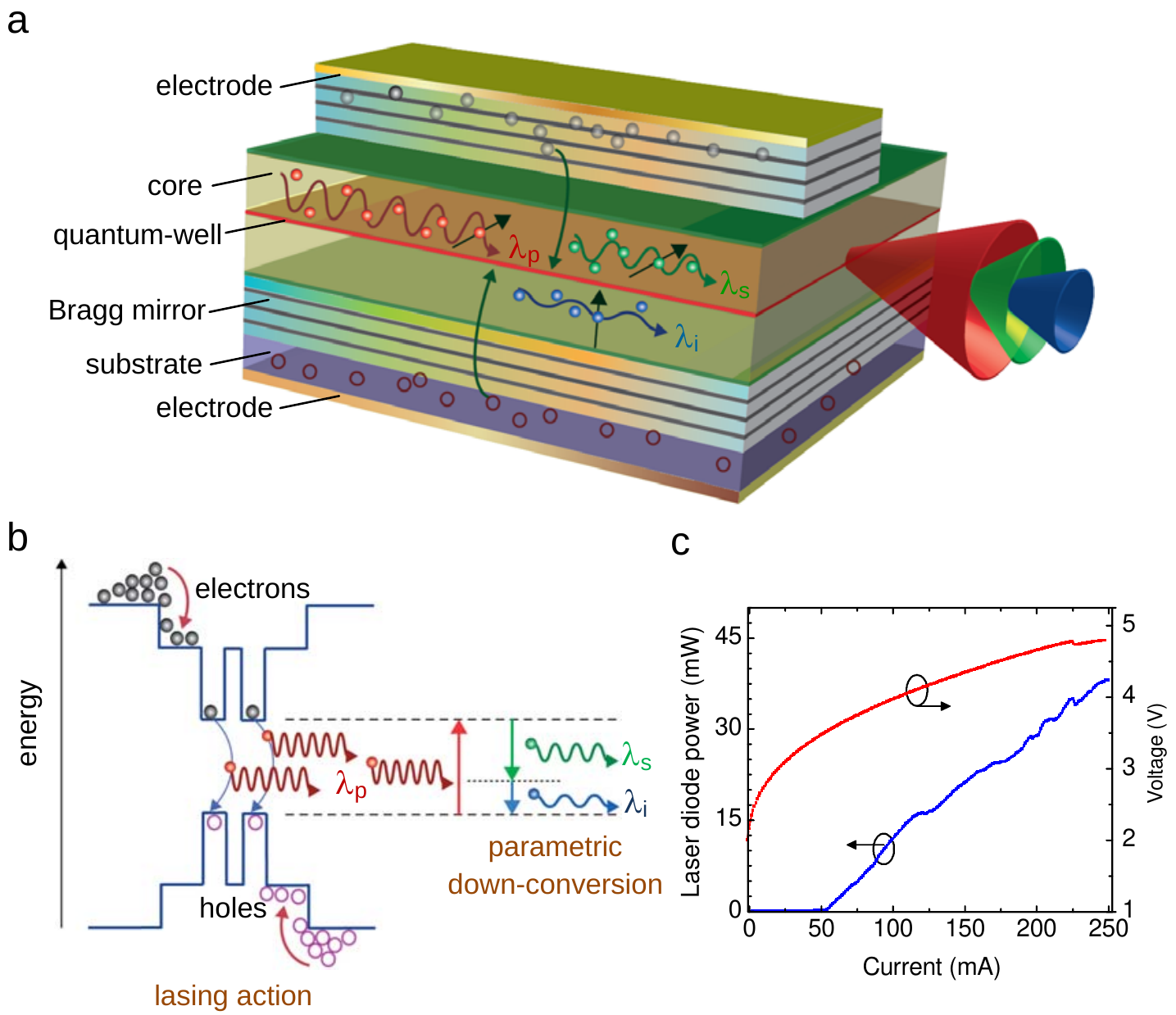}\\
\includegraphics[width=0.6\columnwidth]{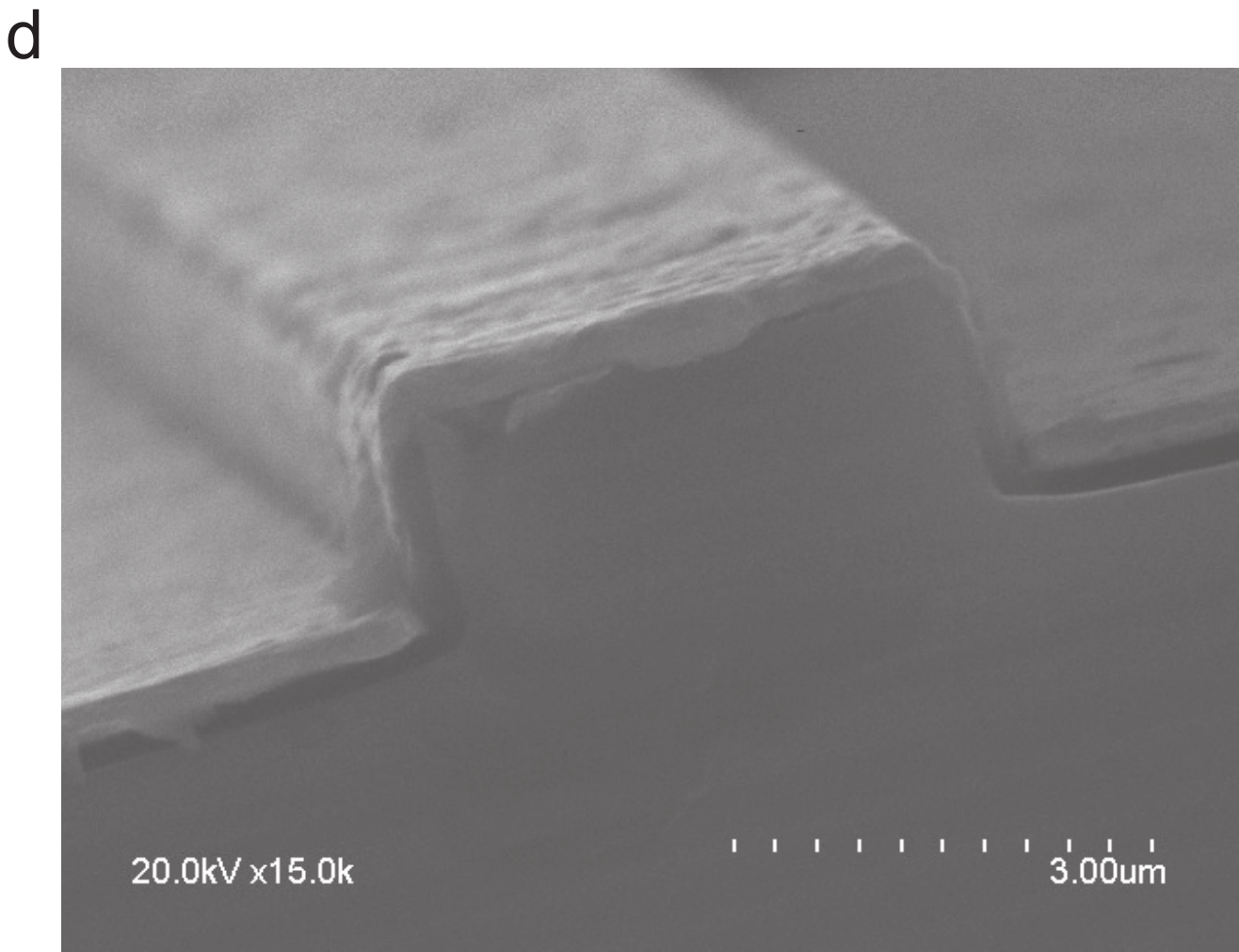}
\caption{Self pumped semiconductor sources for quantum light. (a)the structure of the laser. (b) the energy level diagram of the device elucidating the down conversion process. (c) the light, current, voltage curves of the pump laser. (d)A scanning electron micrograph of the waveguide cross-section.}\label{SAMPLE}
\end{figure}

The semiconductor SPDC waveguide used is based on the AlGaAs material platform. 
The dimension of the waveguide is \(\simeq 1\)mm long and \(\simeq 5\mu m\) wide as shown in Fig. \ref{SAMPLE}.
The ridge and the substrate of the waveguide are based on Bragg structure, which consists of alternating layers of AlGaAs alloy with different compositions (\(Al_xGa_{1-x}As,0\le x \le1\)). 
The semi-periodical structure provides simultaneous confinement of the 783nm pump light and the generated photon pairs around 1566 nm. 
In addition, the waveguide structure is also designed to satisfy the momentum matching condition that is required for the SPDC process to occur. 
The conversion efficiency is estimated to be 2.1\(\times 10^{-8}\) \cite{Horn:2012} for a probe-reference photon pair to be generated by a 783nm pump photon.\\

\subsection{Measured Performance}
\begin{figure}[h]
    \centering
    \includegraphics[width=0.4\columnwidth]{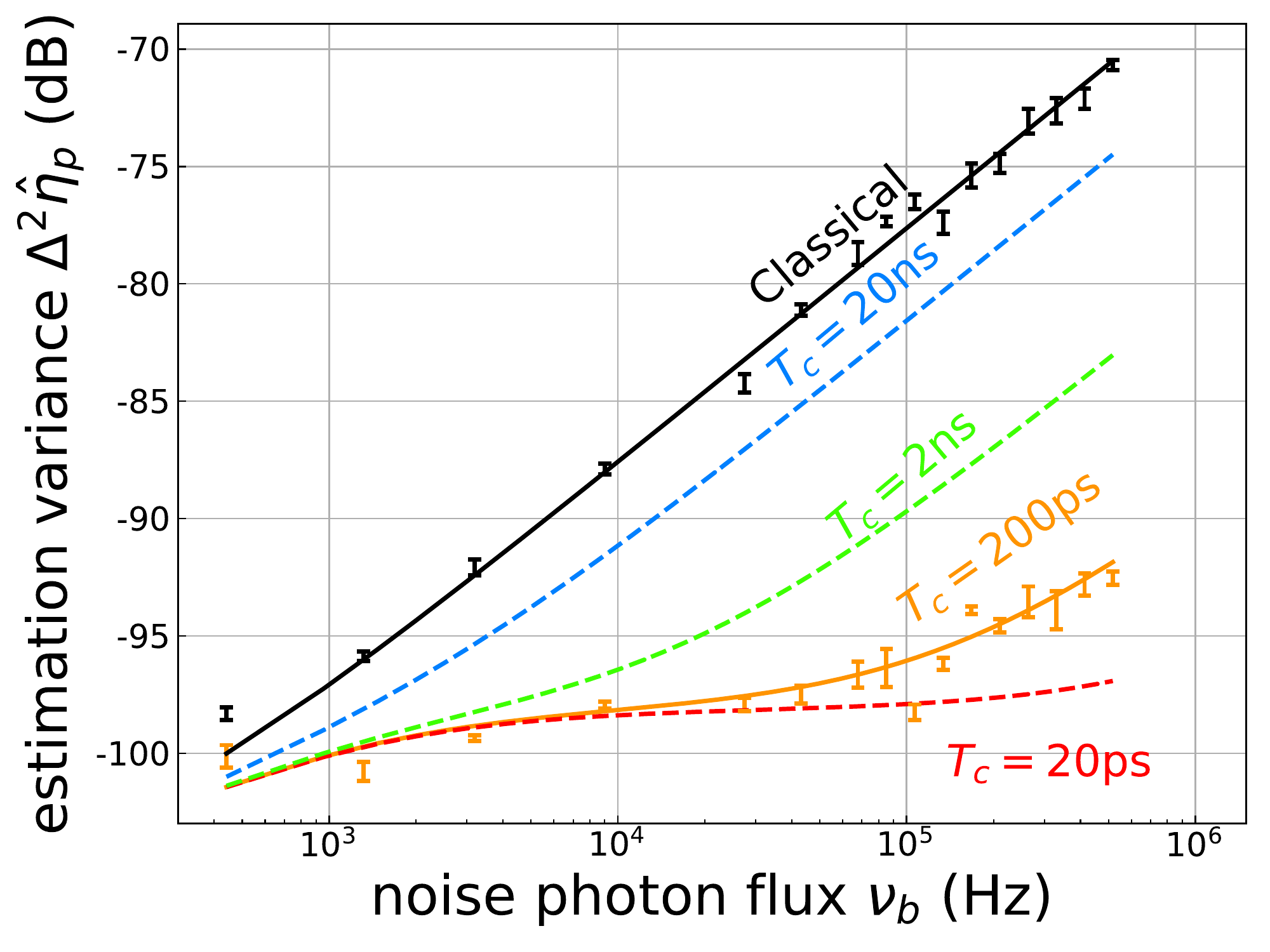}\label{n}  
    \includegraphics[width=0.4\columnwidth]{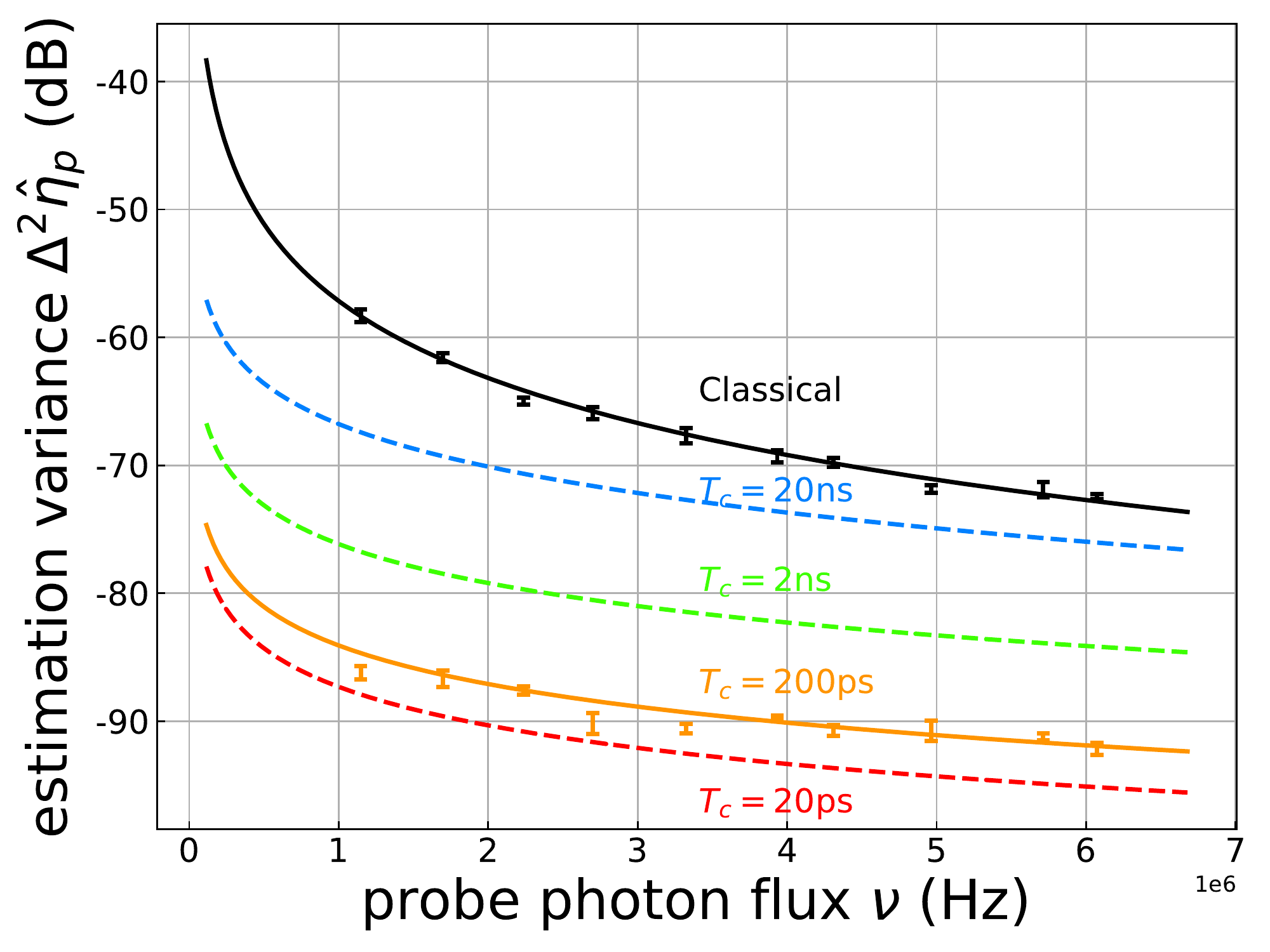}\label{p}   
    \caption{Error bars and solid curves: the experimentally measured and theoretical predicted  variance of the transmission estimation with different environment loss (top) and noise (bottom) condition for both the QTC (orange) and the CI (black) protocol. Dashed curves: theoretically predicted estimation variance for the QTC protocol with different detector temporal resolution, which equals to half of the coincidence window \(T_c\). Details about the experiment result could be found in  \cite{liu2019enhancing} }\label{FI_RESULT}
\end{figure}

We experimentally demonstrated the QTC protocol to confirm its performance enhancement over its classical counterpart (the CI protocol). The target detection experiments of the QTC are carried out under different conditions (probe light flux and noise light flux). The experiment data for the CI protocol is obtained from the experiment data of the QTC protocol by neglecting the photon detection events on the reference detector (since CI could be treated as a special case of QTC with zero reference photon transmission).
By doing so the experimental conditions of the QTC experiment and CI experiment are made sure identical. 
The variance of the transmission estimation as a function of varying probe light flux and noise light flux is shown in Fig. \ref{FI_RESULT}. The experimental result is in close agreement with the theoretical prediction of the Fisher information calculation.  
As can be seen, the performance enhancement of the QTC protocol is most pronounced in the high noise and low probe light flux regime, with the enhancement (defined as the ratio between the estimation variance for the QTC and CI protocol) as a function of noise, and probe light flux reaching up to 21.36 dB, 26.3 dB respectively. In the current implementation of the QTC protocol, one of the major performance limiting factor is the detector temporal resolution \(\simeq 100\)ps. The detector temporal resolution and the target detection performance can be (effectively) improved with better single-photon detection technology or novel photon detection techniques \cite{liu2019enhancing} as shown in Fig. \ref{FI_RESULT}. This is in contrast to the intensity correlation protocol where the correlation enhancement is solely determined by the mean probe photon number per pulse \(\bar{n}\) and limited to the low \(\bar{n}\) regime.   \\

\begin{figure}[h]
\centering
\includegraphics[width=0.4\columnwidth]{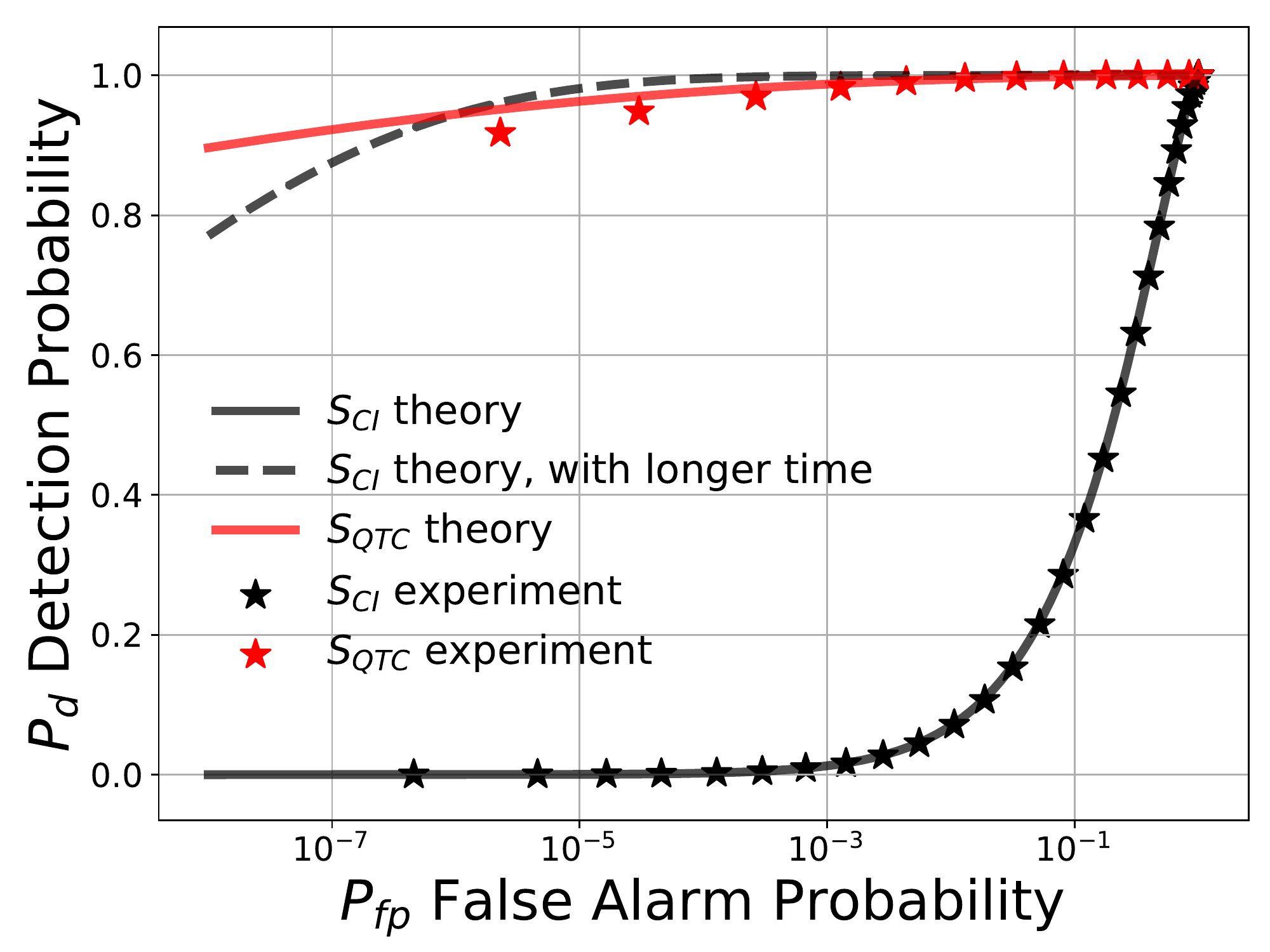}
\includegraphics[width=0.4\columnwidth]{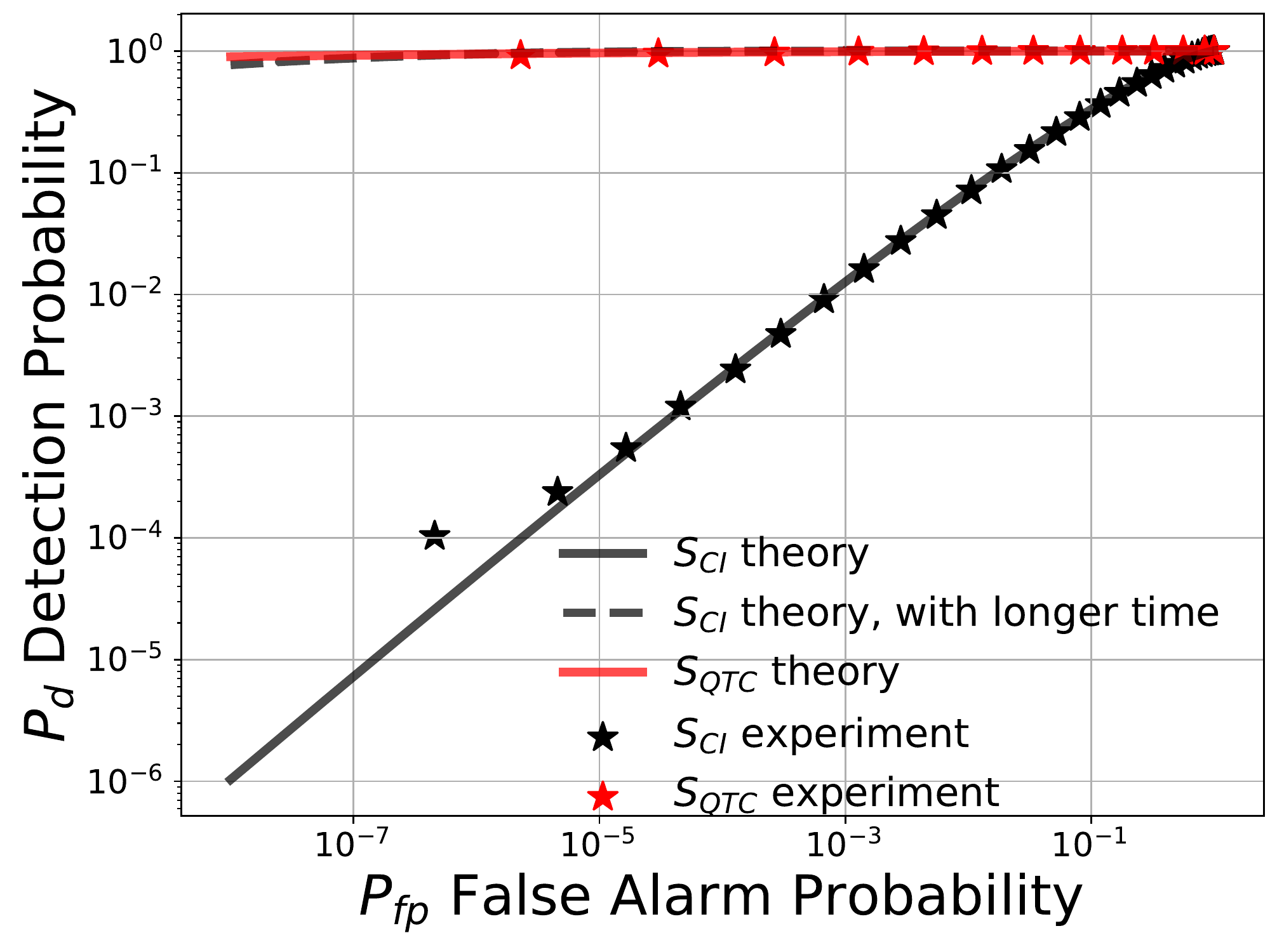}\\
\includegraphics[width=0.4\columnwidth]{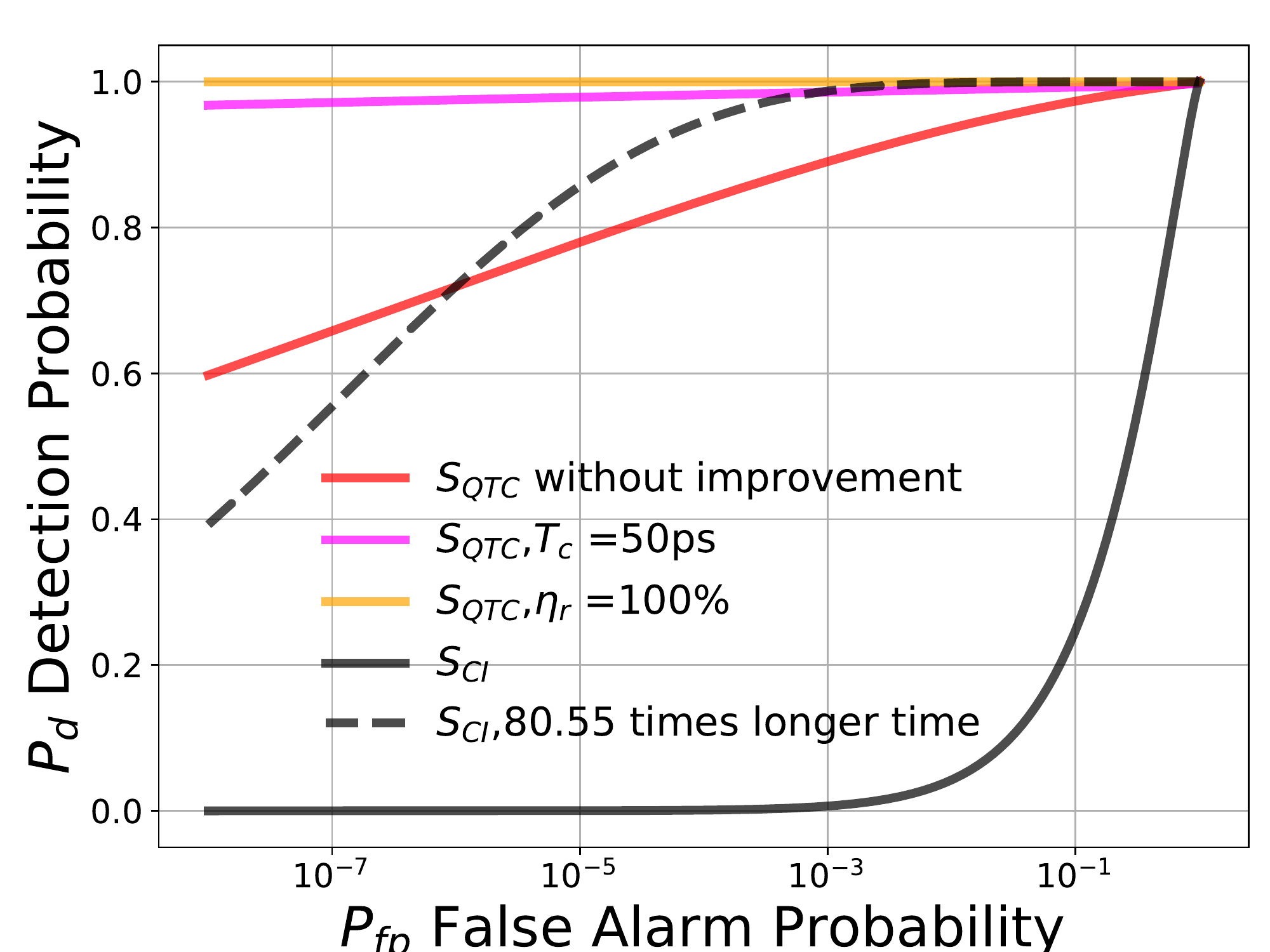}
\caption{Top and middle:the experimentally measured and theoretically predicted ROC curves for the QTC and the CI protocol in linear and log scale. To reach the same detection rate \(P_\text{d}\), with the false alarm rate \(P_\text{fa}=10^{-6}\) and detection time \(\tau=0.01s\), the CI protocol takes 56.75 more detection time. Bottom:the theoretically predicted ROC curve with improved reference channel efficiency and detector temporal resolution. The detection time is set to \(\tau=0.005s\). Details of the experimental parameters could be found in  \cite{trans}. }\label{ROC_FIG}
\end{figure}

The experimentally determined and theoretically calculated ROC curves of the QTC and CI protocol with detection time \(\tau=0.01s\) are shown in Fig. \ref{ROC_FIG}. As can be seen, under the condition of strong environment noise (\(\simeq 20\)dB stronger than the detected probe light power) and at the fixed false alarm rate of \(1\times10^{-6}\), the QTC protocol takes \(\simeq\) 57 times less time to reach the same detection probability \(P_\text{d}\) as the CI protocol. Such detection time reduction suggests that the QTC protocol is capable of detecting the target object with a much higher speed. 
\subsection{Covert Ranging}
\begin{figure}[h!]
\centering
    \includegraphics[width=0.4\columnwidth]{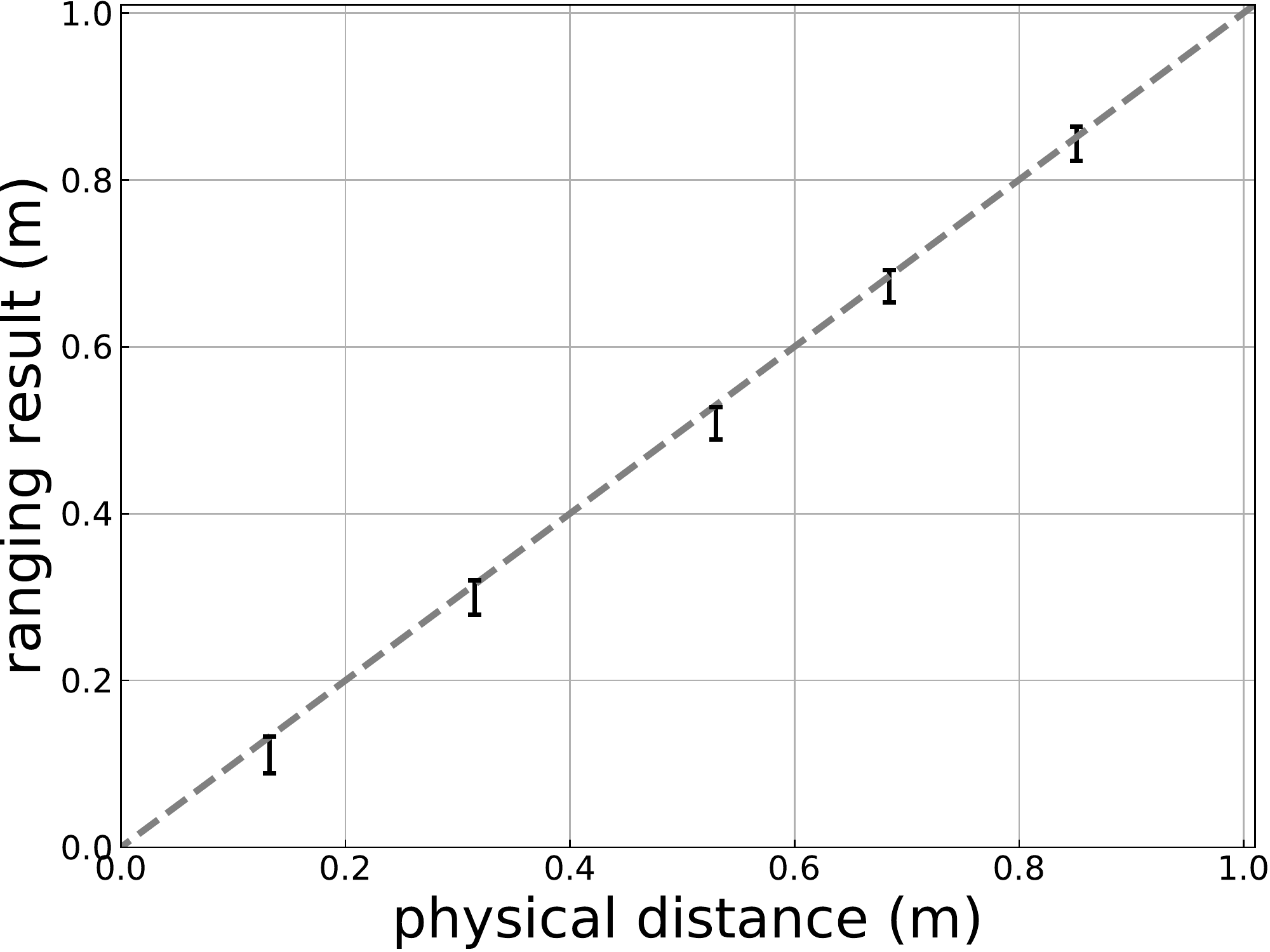}\label{ranging1}
    \includegraphics[width=0.4\columnwidth]{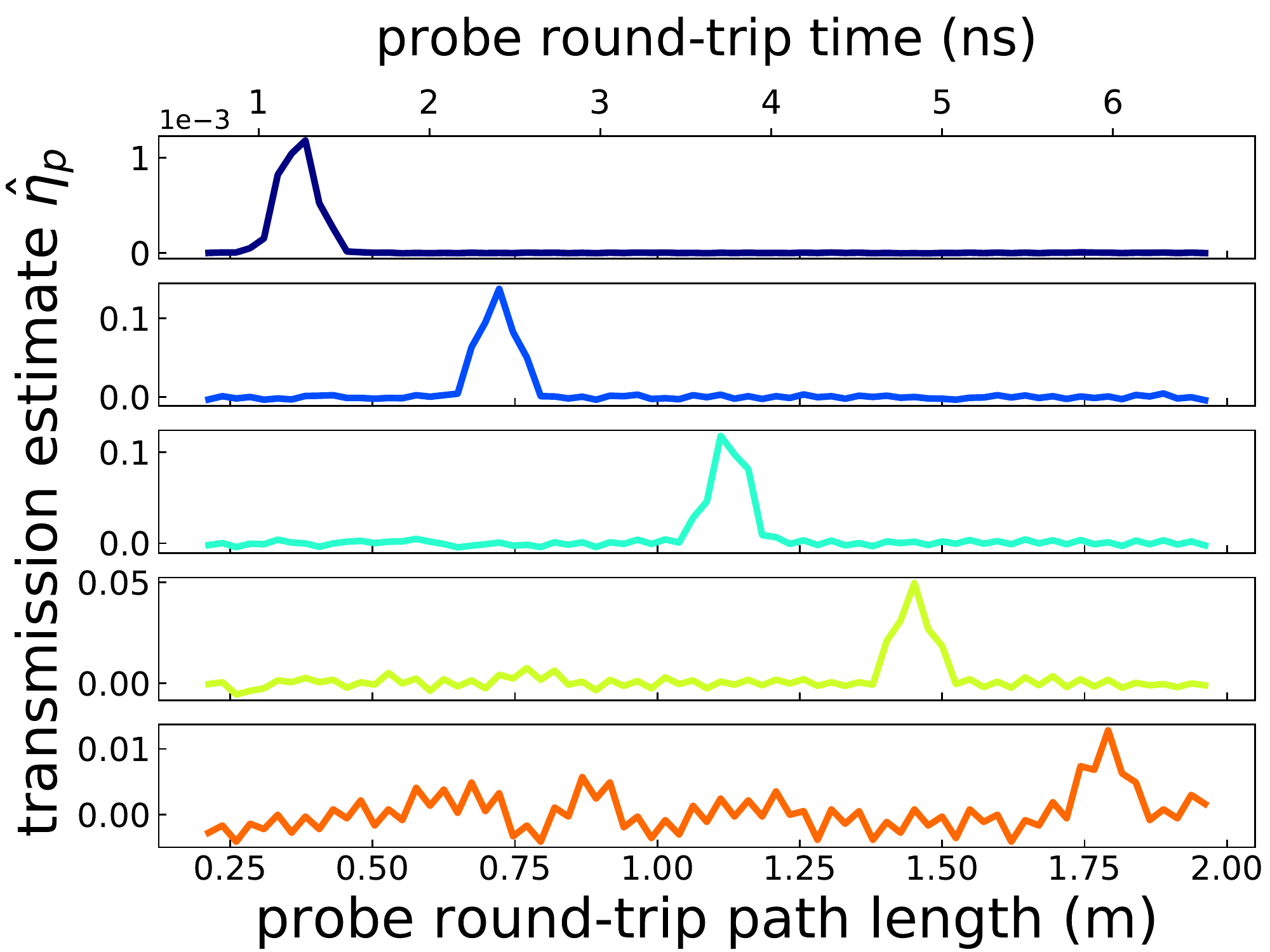}\label{ranging2}
    \caption{Ranging result of the QTC protocol. Top: measured distance versus the physical distance. Bottom: the ``radar signal'' as a function of hypothetical target distance. The 5 panels from top to bottom correspond to physical target distance of 13, 32, 53, 68, 85cm. Details of the experimental parameters can be found in  \cite{liu2019enhancing}}\label{RANGING} 
\end{figure}

In addition to noise resilient target detection, the ability to range with CW probe light is another important application of the QTC protocol.
For classical target detection protocols to determine the distance of the target, pulsed electromagnetic radiation is typically used to probe the target. Then the target distance can be calculated from the time-of-flight of the probe light. 
However, the fact that pulsed light is distinguishable from the CW background noise will make the probe light visible to the unauthorized receivers and therefore vulnerable to active attacks. 
Ranging with classical time-invariant radiation is not possible since the time-invariant back-reflected signal does not contain much information about the target distance. 
On the other hand, ranging with the CW probe light is indeed possible for the QTC protocol. 
This is because each probe photon is temporally correlated with a reference photon, and the travel distance of the probe light can be calculated from the detection time difference of the probe and reference photon. 
Since each probe photon in the QTC detection protocol is generated at fundamentally random time and frequency, they are indistinguishable from the CW broadband environment noise. 
A preliminary covert ranging experiment result has been shown in  Fig. \ref{RANGING}. The maximal distance resolution is around \(\simeq\)5 cm, which is limited by the detector temporal resolution.
\section{DISCUSSION}
\subsection{Technological Possibilities}
The experimental demonstration of the QTC protocol with a semiconductor waveguide source has shown that substantial correlation enhancement of target detection performance is possible, particularly in the low-SNR regime. The results were first presented and discussed in detail in  \cite{liu2019enhancing,trans}. No prior work has demonstrated the correlation enhancement in terms of the ROC metric, common in radar signal processing literature, and at the practically interesting regime of $P_{\text{fa}}=10^{-6}$. 
However, while this is interesting, the question is whether the demonstrated system is favorable in terms of practical implementations. After all, in principle, there are other possible routes to have CW pumped SPDC sources using bulk optics (a combination of lenses/mirrors, etc.) that could reproduce our experimental results. However, while interesting, such sources based on bulk optics are not practical, as they place several demanding requirements, such as optical alignment and difficulties in integration. In particular, the advantages of the semiconductor waveguide photon-pair source for target detection as compared to conventional SPDC sources based on bulk nonlinear crystals are many-fold: 
\begin{itemize}
\item The current experimental demonstration has been using a single transmit module. For practical applications, this is clearly not useful, such as for long-range applications. However, the very small form factor (1mm x 5\(\mu\)m) of the semiconductor waveguide enables large scale integration of photon pairs sources, which is crucial for realistic target detection applications where a large flux of non-classical photon pairs is needed. For example, in the currently used sample, more than 50 waveguides are integrated on a 0.2mm x 1mm x 1cm semiconductor chip. The density of the sources could be further increased with two-dimensional integration. 
  
\item For each of the waveguides, the generation of the probe-reference photon pairs is also efficient because of the strong nonlinearity of the semiconductor material. The AlGaAs platform also allows the possibility of the active waveguide with electrically pumping (i.e. generating the correlated photon pairs by applying a voltage across the waveguide), which is also favorable in terms of large scale integration. 

\item With different designs of the waveguide structure, the temporal spectral property of the SPDC photon pairs could be engineered to improve the QTC protocol performance \cite{Abolghasem:2009}. For example, with an appropriately designed waveguide structure, the generated photon pairs could have an extremely strong intrinsic temporal correlation \(\Delta t\simeq 20fs\) \cite{Abolghasem:2009}, which translates to high noise resilience of the QTC protocol. Furthermore, the central frequency of the probe light could be shifted (from IR to THz) to suit the need in different target detection scenarios without shifting the central frequency of the reference photons.  
\end{itemize}

\subsection{Performance Limiting Factors}
For the current implementation of the QTC protocol, an important performance-limiting factor is the transmission efficiency of the reference photons \(\eta_r\simeq 20\%\), which could be due to the inefficient coupling of the reference photons from the waveguide to the single-mode fiber that is connected to the detector. 
Such inefficiency could be alleviated with improved optical setup.
In the limit of perfect reference photon transmission (\(\eta_r = 100\%\)) the ROC curve of the QTC detection protocol is plotted in Figure \ref{ROC_FIG}. 
Another important limiting factor of the QTC protocol performance is the detector temporal resolution \(\Delta t_{det}\simeq 121\)ps, which limits the maximal amount of intrinsic temporal correlation \(\Delta t\) of the photon pairs that could be utilized for target detection. 
Such a limit could be overcome with better single-photon detection technologies or with improved photon detection techniques\cite{liu2019enhancing}. Fig. \ref{ROC_FIG} also shows the ROC curve of the QTC detection protocol when the detector temporal resolution is improved to \(\Delta t_{det}\simeq 12\) ps.\\

Factors affecting both the QTC protocol and the CI protocol performances include the source power, background noise power and the probe photon transmission efficiency \(\eta_p\). However, it is worth noting that the noise power does not affect the QTC and the CI detection protocols equally. The performance advantage of the QTC detection protocol over the CI detection protocol is most pronounced with a high level of background noise, suggesting that the QTC protocol is more favorable for target detection with a high noise background. The total probe transmission efficiency, \(\eta_p\), impacts the performance of both the QTC detection protocol and the CI detection protocol. The total transmission \(\eta_p\) is affected by many factors including the target distance and cross-section, the reflectivity of the target, propagation medium, and the probe photon collection efficiency.

\subsection{Covert Operation Of The QTC Protocol}
The experimental result above shows the significant performance advantages of the QTC detection protocol over the CI detection protocol. However, it is important to note that for both the QTC and CI protocol, the background noise is assumed to be overlapping with the probe photon in both temporal and spectral domains, i.e., broadband and CW. Therefore it is not possible to reduce the in-band noise power through filtering. 
If narrowband or pulsed probe light is to be used for the CI protocol, it is indeed possible to increase its performance through spectral filtering or temporal gating to reduce the in-band noise power. 
However, it must also be noted that such noise reduction requires concentrated optical power in either temporal or spectral domain, which increases the visibility and vulnerability of the target detection channel. 
For example, if the adversary parties is able to distinguish the probe photons from the background noise by their different spectral-temporal property, then they could selectively jam the target detection channel with noise photons that are indistinguishable from the probe photons. 
Although there exist classical scrambling techniques such as frequency scrambling to increase the indistinguishability between the probe and noise photon, such indistinguishability is not guaranteed by fundamental physical principles. 
On the other hand, the QTC detection protocol provides unconditional indistinguishability between the probe and the noise photons. 
This is because each of the probe photons is generated at a fundamentally random time and frequency, albeit its strong intrinsic temporal correlation to the reference photon. 
When the environment noise power is high compared to the probe power, the probe photon will be almost invisible to the adversary, but the QTC detection protocol could still achieve nontrivial target detection performance, as shown in the results and analysis above.

\section{CONCLUSION}
In this article, we discussed how one can exploit the correlations within non-classical light sources to enhance the target detection performance, in a promising approach.
In particular, we showed how the QTC protocol can exploit the strong temporal correlation of non-classical photon pairs to substantially enhance the target detection performance in a noisy and lossy environment.  
The QTC protocol uses probe photons that are completely indistinguishable from the background noise for unauthorized receivers, and could be used for covert ranging and target detection. 
The semiconductor waveguide source that we used for the QTC protocol is suitable for practical implementation due to its potential for high density integration, high conversion efficiency, and tunable spectral properties. 
\bibliographystyle{IEEEtran}
\bibliography{./References}
\section*{figure captions}
Fig. 1:  The schematic of general correlation enhanced target detection protocols. Left: the non-classical photon pair source; middle: the free-space transceiver of the probe photons; right: the detection of the probe and the reference photons and the data analysis.\\
      
Fig. 2:  Comparison between the normalized mutual information for target detection protocols enhanced by (blue) intensity correlation and (red,magenta and orange) temporal correlation. The pump pulse duration \(\sigma_+\) for the temporal correlation protocol is set to 1 ns.\\
      
Fig. 3:  The performance advantages (quantified as the ratio of the \(\text{SNR}_{COV}\)) of the intensity correlation protocol as compared to the classical intensity correlation protocol. Blue (teal) line: with (without) the additional loss and noise in the target detection channel. The detail of the experimental result and analysis could be found in  \cite{He:2018}. \\
      
Fig. 4:  Self pumped semiconductor sources for quantum light. (a)the structure of the laser. (b) the energy level diagram of the device elucidating the down conversion process. (c) the light, current, voltage curves of the pump laser. (d)A scanning electron micrograph of the waveguide cross-section.\\
      
Fig. 5:  Error bars and solid curves: the experimentally measured and theoretical predicted  variance of the transmission estimation with different environment loss (top) and noise (bottom) condition for both the QTC (orange) and the CI (black) protocol. Dashed curves: theoretically predicted estimation variance for the QTC protocol with different detector temporal resolution, which equals to half of the coincidence window \(T_c\). Details about the experiment result could be found in  \cite{liu2019enhancing} \\
      
Fig. 6:  Top and middle:the experimentally measured and theoretically predicted ROC curves for the QTC and the CI protocol in linear and log scale. To reach the same detection rate \(P_\text{d}\), with the false alarm rate \(P_\text{fa}=10^{-6}\) and detection time \(\tau=0.01s\), the CI protocol takes 56.75 more detection time. Bottom:the theoretically predicted ROC curve with improved reference channel efficiency and detector temporal resolution. The detection time is set to \(\tau=0.005s\). Details of the experimental parameters could be found in  \cite{trans}. \\
      
Fig. 7:  Ranging result of the QTC protocol. Top: measured distance versus the physical distance. Bottom: the ``radar signal'' as a function of hypothetical target distance. The 5 panels from top to bottom correspond to physical target distance of 13, 32, 53, 68, 85cm. Details of the experimental parameters can be found in  \cite{liu2019enhancing}\\

\end{document}